\documentclass[pra,twocolumn,longbibliography]{revtex4-1}
\usepackage[utf8]{inputenc}
\usepackage{amsmath}
\usepackage{amsfonts}
\usepackage{amssymb}
\usepackage{bm}
\usepackage{graphicx}
\usepackage{xcolor}
\usepackage{microtype}

\begin{document}
\title{Microscopic characteristics and tomography scheme of the local Chern marker}
\author{Bernhard Irsigler, Jun-Hui Zheng, and Walter Hofstetter}
\affiliation{Institut f\"ur Theoretische Physik, Goethe-Universit\"at Frankfurt am Main, Germany}
\begin{abstract}
The concept of the local Chern marker has gained a lot of attention especially in the field of ultracold quantum gases in optical lattices and artificial gauge fields. We investigate in further detail the microscopic real-space characteristics of the local Chern marker for the two-band Harper-Hofstadter-Hatsugai model and propose a tomographic scheme for the experimental detection of an approximate local Chern marker neglecting higher orders.
\end{abstract}
\maketitle
\section{Introduction}
Non-trivial topological invariants are the fundamental reason for robust edge states. The vast diversity of different topological invariants opens up a whole world of topological insulators and topological superconductors \cite{Hasan2010,Wen2017,Hofstetter2018,Rachel2018}. The deep connection between topology and physics originally manifests in the TKNN invariant \cite{Thouless1982} which corresponds to the first Chern number. It is defined as a $k$-space integral of the Berry curvature and thus solely a bulk property. For the case of disorder real-space versions of the Chern number have been successfully applied \cite{Prodan2010,Petrescu2012}. A global topological invariant based on real-space representation is the Bott index \cite{Loring2010}. Bianco and Resta \cite{Bianco2011}, however, derived a real-space expression by Fourier transforming the Chern number and omitting a final real-space integration. The obtained position-resolved quantity is called local Chern marker. Since its introduction it enjoys great popularity in condensed matter theory as well as in the field of cold atomic gases. Applications range from heterojunctions \cite{Bianco2011} and quasicrystals \cite{Tran2015} to interacting fermions \cite{Amaricci2017} and interacting, spin-orbit coupled fermions at the smooth topological interface \cite{Irsigler2019} and trap geometries \cite{Gebert2019}. 

In the last years enormous progress has been made implementing topological phases in cold atom experiments \cite{Aidelsburger2013,Miyake2013,Jotzu2014}. In these clean und highly controllable setups it was possible to measure the Chern number \cite{Aidelsburger2014} as well as probing the Berry curvature \cite{Flaschner2016}. In contrast to solid state materials, cold atom setups are intrinsically inhomogeneous due to confining laser potentials which makes local topological invariants especially interesting for cold atomic gases. A recent theoretical study of non-interacting, non-equilibrium dynamics showed an intriguing current of the local Chern marker \cite{Caio2019}. In our work, we investigate the microscopic characteristics of the local Chern marker by analyzing its contributions and propose an approximate experimental implementation for cold atom setups for the contributions. A recent experiment in a photonic implementation measured the Chern number locally \cite{Kitaev2006,Mitchell2018} from electromagnetic response \cite{Schine2019}. 

The paper is structured as follows. We introduce the local Chern marker in Sec.~II and the Harper-Hofstadter-Hatsugai (HHH) model in Sec.~III. In Sec.~IV we discuss the contributions to the local Chern marker explicitly for the HHH model and its finite size scaling. In Secs.~V and VI we apply our theory to the harmonically trapped system and a topological interface, respectively, as examples for inhomogeneous cold atom setups. In Sec.~VII we introduce a tomography scheme to measure the most prominent contribution to the local Chern marker. In Sec.~VIII we conclude.

\section{Local Chern marker}
The local Chern marker is introduced in Ref.~\cite{Bianco2011} by rewriting the expression for the Chern number in real-space representation, i.e., with the substitution $\langle u_{n'\bm{k}}|\partial_{\bm{k}}| u_{n\bm{k}}\rangle=-i\langle\psi_{n'\bm{k}}|\bm{r}|\psi_{n\bm{k}}\rangle$. Here, $|\psi_{n\bm{k}}\rangle = e^{i\bm{r}\bm{k}}|u_{n\bm{k}}\rangle$ is a Bloch state with quasimomentum $\bm{k}$ in band $n$, $|u_{n\bm{k}}\rangle$ is the periodic part of the Bloch state, and $\bm{r}=(x,y)$ denotes a position vector. One way to write the local Chern marker is
\begin{equation}
C(\bm{r}) = -4\pi \mathrm{Im}\langle\bm{r}|\hat{P}\hat{x}\hat{P}\hat{y}\hat{P}|\bm{r}\rangle,
\end{equation}
where $\hat{P}=\sum_{n\in O}\int d\bm{k}|\psi_{n\bm{k}}\rangle\langle\psi_{n\bm{k}}|$ is the projection operator onto the set of occupied states $O$. Furthermore, $\hat{x}$ ($\hat{y}$) is the $x$ ($y$) component of the position operator. The local Chern marker can be rewritten in terms of the single-particle density matrix $\rho(\bm{r},\bm{r'})=\sum_{n\in O}\psi_n^*(\bm{r})\psi_n(\bm{r'})=\langle \bm{r'}|\hat{P}|\bm{r}\rangle$, where $|\bm{r}\rangle$ are eigenstates of the position operator:

\begin{equation}
C(\bm{r})=4\pi\mathrm{Im}\sum_{\bm{r'},\bm{r''}}\rho(\bm{r},\bm{r'})x'\rho(\bm{r'},\bm{r''})y''\rho(\bm{r''},\bm{r}).
\label{LCM}
\end{equation}

This expression, as it is derived from the Chern number, is gauge invariant. Contributions, where any pair of $\bm{r}$, $\bm{r'}$ or $\bm{r''}$ is equal, vanish since they are purely real, e.g., if $\bm{r}\neq\bm{r'}=\bm{r''}$, we find $|\rho(\bm{r},\bm{r'})|^2n(\bm{r'})x'y'$, where $n(\bm{r})=\rho(\bm{r},\bm{r})$ is the on-site density.

The three corner points lead us to the notion of triangles. The local Chern marker is thus a sum of contributions from all possible triangles $(\bm{r},\bm{r'},\bm{r''})$ of off-diagonal density matrices with one corner at lattice site $\bm{r}$. 

In contrast to its name, the local Chern marker is rather quasi-local. This can be observed in Eq.~\eqref{LCM} which contains two sums over lattice positions which span the whole lattice. However, if the system is not close to a topological phase transition the contributions to the local Chern marker are sufficiently local.

\begin{figure*}
\centering
\includegraphics[width=\textwidth]{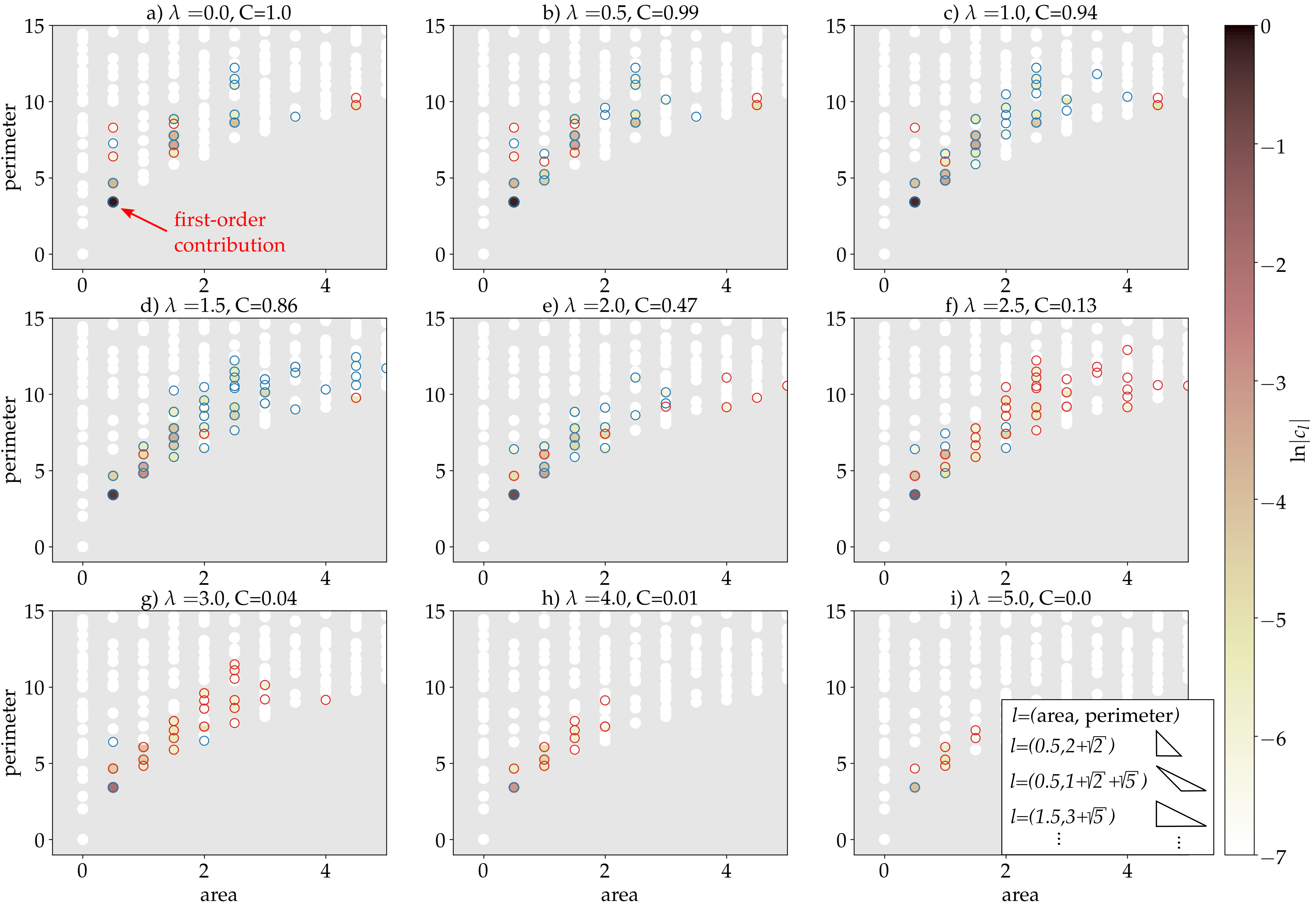}
\caption{Contributions $c_l$ to the local Chern marker as function of the triangle's area and perimeter in color for different values of the staggering potential $\lambda$. The legend depicts examples for the triangles which are summed in each contribution. The  positive (negative) sign of each contribution is marked in the figure by a blue (red) circle around the data point. $C$ refers to the bulk value of the local Chern marker. Results were obtained for a $40\times40$ lattice. We introduce a lower cutoff for the contributions of $10^{-7}$ for better visibility. We highlight the first-order contribution in a).}
\label{contr}
\end{figure*}

\section{Harper-Hofstadter-Hatsugai model}
To study the contributions explicitly we focus on the HHH model \cite{Hatsugai1990} which exhibits a gapped topologically non-trivial phase at half filling, in contrast to the Hofstadter model \cite{Aidelsburger2013}, and has some similarity to the Haldane model \cite{Jotzu2014}.
The Hamiltonian of the HHH model reads
\begin{equation*}
\begin{split}
\hat{H}&= \lambda\sum_{\bm{r}}(-1)^x\hat{c}_{\bm{r}}^\dag\hat{c}_{\bm{r}}-t\sum_{\bm{r}}\left[
\hat{c}_{\bm{r}+\vec{x}}^\dag\hat{c}_{\bm{r}}
+e^{2\pi i \alpha x}\hat{c}_{\bm{r}+\vec{y}}^\dag\hat{c}_{\bm{r}}\right.\\
&\left.+e^{2\pi i \alpha (x+1/2)}\hat{c}_{\bm{r}+\vec{x}+\vec{y}}^\dag\hat{c}_{\bm{r}}
+e^{2\pi i \alpha (x+1/2)}\hat{c}_{\bm{r}+\vec{y}}^\dag\hat{c}_{\bm{r}+\vec{x}}
+\text{h.c.}\right]
\end{split}
\end{equation*}
where $t$ is the hopping energy and serves as our energy scale in the following, i.e.,  $t=1$,  $\vec{x}$ $(\vec{y})$ denotes unit vectors in the $x$ $(y)$ direction, $\alpha$ is the square plaquette flux, and $\lambda$ is an additional staggering potential. We focus on the case $\alpha=1/2$ since it yields a two-component unit cell and thus a two-band model. We label the two components of the unit cell by $A$ and $B$ as shown in Fig.~\ref{latt}a) and restrict ourselves to the half-filled case. The Fourier-transformed Hamiltonian reads $\hat{H}(\bm{k})=v(\bm{k})\cdot\bm{\sigma}$, where $v(\bm{k})=\left(2\cos(k_x),4\sin(k_x)\sin(k_y),\lambda-2\cos(k_y)\right)$ \cite{Zheng2018a}. Here, $\bm{\sigma}$ refers to the Pauli vector in the pseudospin representation of the ($A$,$B$) basis. The system exhibits a topological phase transition at the critical staggered potential $\lambda_c=2$. For a two-level system, this can be determined by whether the two-dimensional surface of $v(\bm{k})$ encloses the origin $\bm{k}=0$. This is equivalent to the notion of a two-dimensional generalization of a winding number (see \cite{Hasan2010}, Sec.~II.B.2).
For $|\lambda|<2$, $v(\bm{k})$ encloses the origin and the system is in a topologically non-trivial phase with Chern number 1. For $|\lambda|>2$, $v(\bm{k})$ does not enclose the origin and the system is in a topologically trivial phase with Chern number 0.

\section{Contributions to the local Chern marker}
As discussed in Sec.~II the local Chern marker is a sum of off-diagonal density matrices connected to triangles. If we characterize the triangle by its perimeter and its area combined in a tuple $l=$(area, perimeter) the summation can be rewritten as $C(\bm{r})=\sum_lc_l$. Here, $c_l$ is the sum of contributions of triangles having the same perimeter and area $l$, examples are given in Fig.~\ref{contr}. 
Fig.~\ref{contr} shows $c_l$ of the first few triangles as function of their area and perimeter for different staggering potentials $\lambda$ in a)-i) computed in a $40\times40$ lattice. Triangles with vanishing area do not contribute to the local Chern marker. 

We observe that the contribution of the triangles with area 0.5 and perimeter $2+\sqrt{2}$ is the largest which is highlighted in Fig.~\ref{contr}a). In the following, we call it first-order contribution. For $\lambda=0$ in Fig.~\ref{contr}a) it exceeds by three orders of magnitude the second largest contribution and amounts to roughly 0.87, i.e., 87\% of the quantized value of 1. The missing 13\% seem to stem from long-range correlations over the lattice. Note that this value is model-specific. The bulk average of the local Chern marker $C$ is computed as the ($A,B$) average in the center of the system $C=\left[C(\bm{r}_A)+C(\bm{r}_B)\right]/2$. Since we consider a two-band model, the average has two contributions. For $\lambda=0$ it is quantized and corresponds to the Chern number. For increasing $\lambda$, we observe that it is not quantized anymore. Also the first-order contribution decreases and simultaneously more positive-valued (marked by a blue circle) higher-order contributions emerge. This is best seen right before the phase transition in Fig.~\ref{contr}d) for $\lambda=1.5$. Right after the phase transition for $\lambda=2.5$, in Fig.~\ref{contr}f), we observe roughly as many higher-order contributions as in Fig.~\ref{contr}d) but with a negative sign (marked by a red circle). We interpret these results as long-range correlations over the whole lattice near the phase transition due to the gap closing. This is why finite-size effects will always emerge and the local Chern marker is not quantized close to the phase transition $\lambda=\lambda_c$.

\begin{figure}
\centering
\includegraphics[width=\columnwidth]{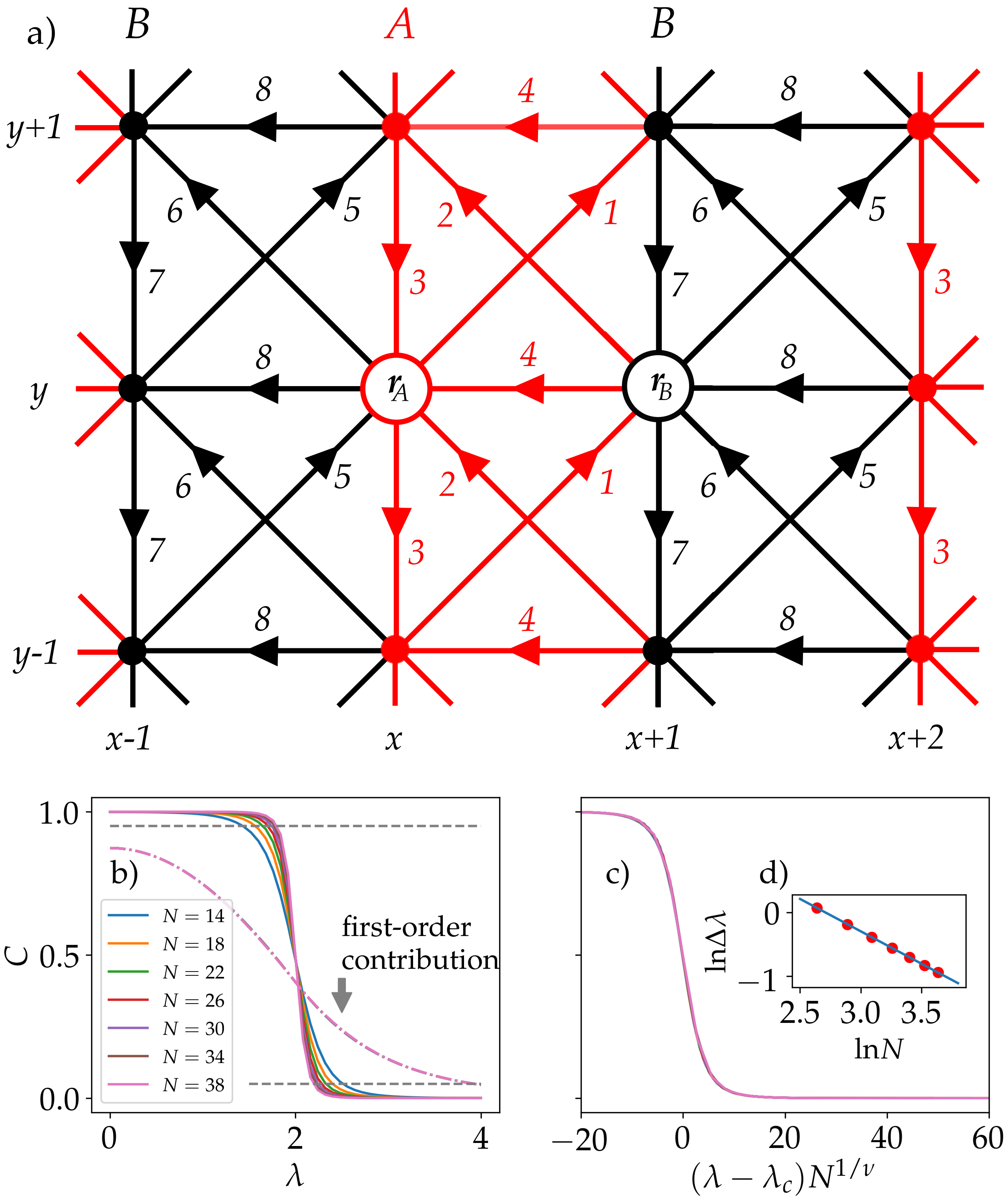}
\caption{a) Labeling of the density matrices for the first-order contribution to the local Chern marker in the ($A,B$) two-level model. Off-diagonal density matrices are uniquely labeled by a number and a direction shown as an arrow, and are connected to triangles. b) finite-size scaling of the bulk average of the local Chern marker, c) collapse of curves after rescaling, and d) power law relation between the width of the transition region $\Delta\lambda$ and the system size $N$.}
\label{latt}
\end{figure}

We now focus on the described first-order contribution and derive a simple expression for the local Chern marker for the two-band HHH model. The first-order contribution only contains terms with density matrices connecting nearest-neighboring and next-nearest-neighboring sites. This already contains contributions of 48 different triangles. We exploit the symmetries of the model in order to reduce this number. In Fig.~\ref{latt}a) we show schematically all possible triangles for the first-order contribution to the local Chern marker. The numbering as well as the arrow direction label a specific density matrix, e.g.,
$\rho_1=\rho(\bm{r}_A,\bm{r}_A+\vec{x}+\vec{y})$. We assume local translational invariance in this small region. Evaluating Eq.~\eqref{LCM} for the triangles in Fig.~\ref{latt} and performing the ($A,B$) average $C$ we find
\begin{equation}
C=-4\pi(\rho_7-\rho_3)(2\rho_5\rho_8+2\rho_6\rho_8^*-\rho_1\rho_4-\rho_2\rho_4^*).
\label{firstOrd}
\end{equation}
The first-order contribution is thus reduced from a sum of 48 different triangles to a formula containing just eight off-diagonal density matrices.

In Fig.~\ref{latt}b) we look at finite size scaling of the two-band HHH model. We show the bulk average $C$ of the local Chern marker for different sizes of the lattice $N\times N$ as a function of $\lambda$ and observe that the transition at $\lambda=\lambda_c$ becomes steeper with increasing lattice size. We also show the first-order contribution as dash-dotted lines and oberserve that it is scale invariant showing that it is purely local.  In Fig.~\ref{latt}c) and d) we perform a scaling analysis of the two-band HHH model according to the scaling analysis of the Haldane model in Ref.~\cite{Caio2019}. To this end, we define the width of the transition region $\Delta\lambda$ as the difference of the value of $\lambda$ where $C=0.05$ and the value of $\lambda$ where $C=0.95$ both represented as two horizontal dashed gray lines in Fig.~\ref{latt}b). We further assume that the bulk correlation length $\xi$ scales as $\xi\approx(\Delta\lambda)^{-\nu}$. Since $\xi$ directly scales width the system size, we find $\Delta\approx N^{-1/\nu}$. From Fig.~\ref{latt}d) we compute $\nu\approx1.02$. We assume the scaling form of the bulk value of the local Chern marker $C\sim f(\xi/N)$ which with the considerations made before becomes $C\sim\tilde{f}\left((\lambda-\lambda_c)N^{1/\nu}\right)$. After the rescaling we observe a collapse of curves in Fig.~\ref{latt}c) like in Ref.~\cite{Caio2019}. In contrast to the result for the Haldane model \cite{Caio2019}, we observe a rescaled curve in Fig.~\ref{latt}c) for the HHH model exhibiting the symmetry $\tilde{f}\left(-(\lambda-\lambda_c)N^{1/\nu}\right)=1-\tilde{f}\left((\lambda-\lambda_c)N^{1/\nu}\right)$.

\begin{figure}
\centering
\includegraphics[width=\columnwidth]{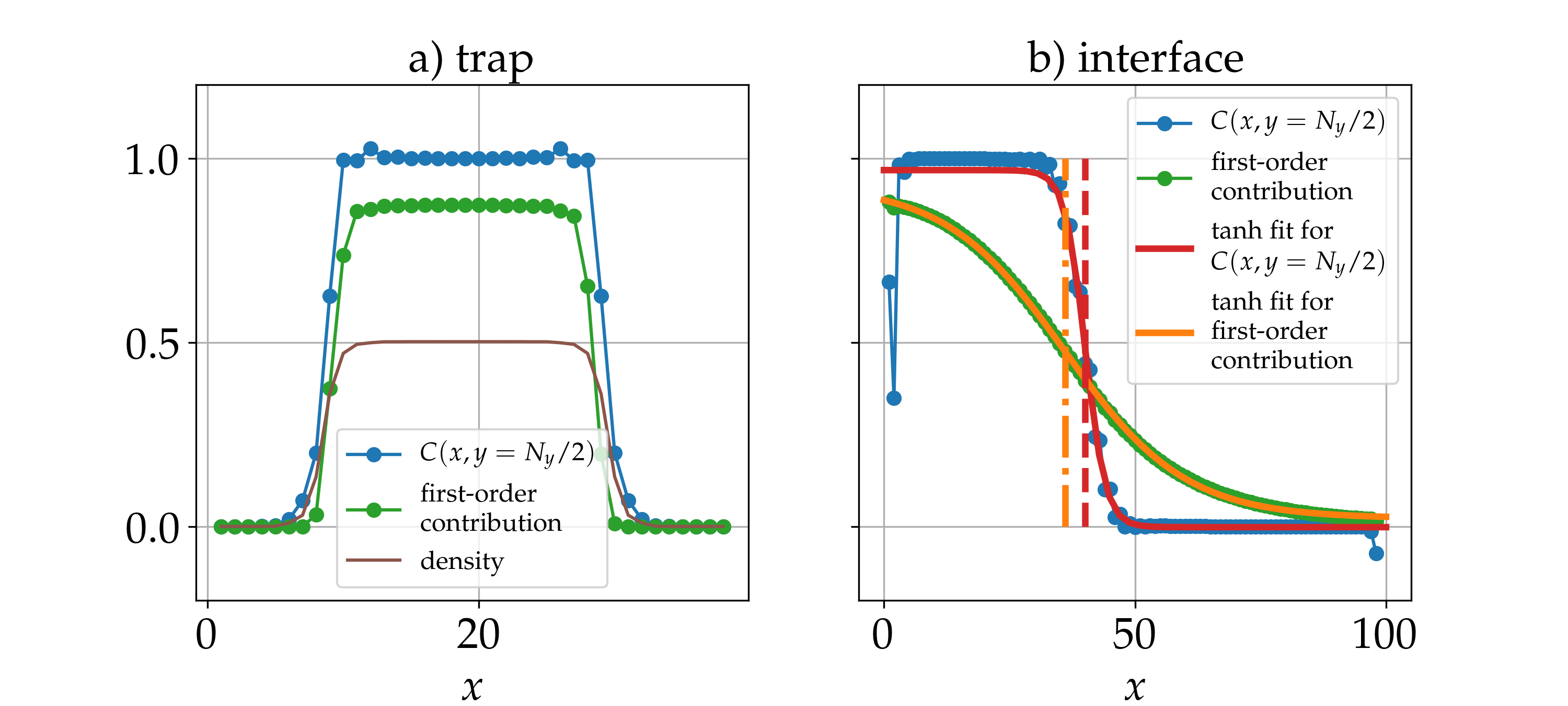}
\caption{Local Chern marker and first-order contribution according to Eq.~\eqref{firstOrd} in inhomogeneous systems: a) the harmonically trapped system and b) the topological interface.}
\label{trap}
\end{figure}

\section{Trap}
As it is a common example for an inhomogeneous system in cold atom setups we check our theory on a system with a harmonic trap. Fig.~\ref{trap}a) shows the local Chern marker as well as the first-order contribution according to Eq.~\eqref{firstOrd}. We observe good qualitative agreement between both approaches. The bulk value deviates as discussed before by about 13\%. 

\section{Interface}
A second example is the topological interface \cite{Reichl2014,Goldman2016,Irsigler2019} which is used to create an in-situ topological phase separation. In Fig.~\ref{trap}b) we show the local Chern marker resulting from a system with staggering potential \mbox{$\lambda(x) = 5x/100(-1)^x$} on a 100$\times20$ lattice. We apply a hyperbolic tangent fit \mbox{$a-b\tanh(c(\lambda-d))$}, where we assume that the phase transition point is given by $d$. The local Chern marker predicts a phase transition point at $x\approx40$, shown as dashed red line, which translates to a critical staggering potential of 2 as expected. The first-order contribution is rather smooth and predicts a phase transition point of $x\approx36$, shown as dash-dotted yellow line, which translates to a critical staggering potential of 1.8. The first-order contribution can thus estimate the phase transition point up to an error of 10\%. On the other hand, in Fig.~\ref{latt}b), we showed that the first-order contribution is scale invariant and always takes the value 0.4 at the phase transition. With this consideration, the phase transition point is determined as $x\approx40$, corresponding to a critical staggering potential of 2 as expected. Of course this value is also model dependent.

\begin{figure}[ht]
\centering
\includegraphics[width=\columnwidth]{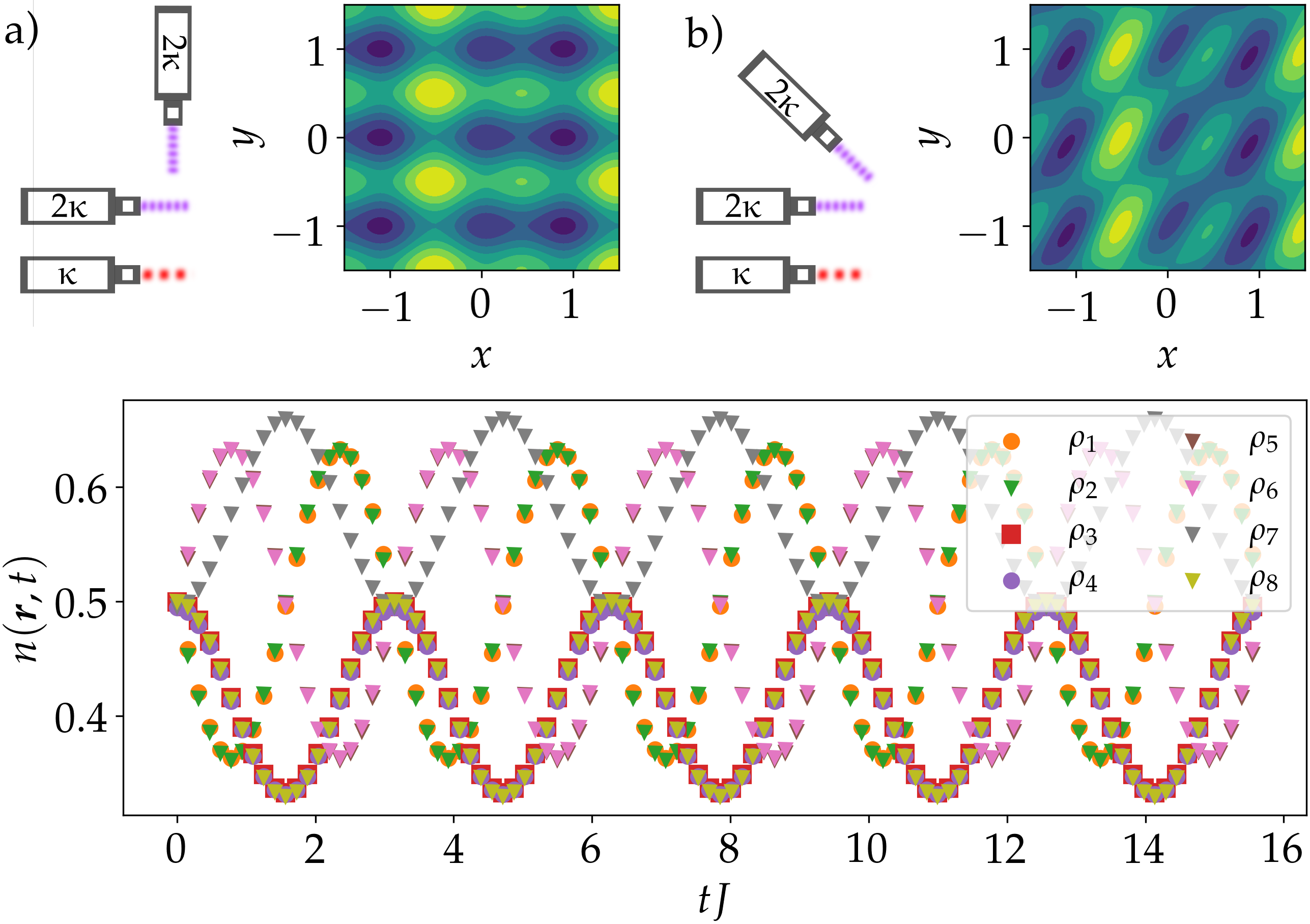}
\caption{Quench Hamiltonians realized by superlattice structures for a) nearest-neighbor and b) next-nearest-neighbor couplings. Dark regions correspond to minima at the lattice sites. c) Time evolution of the local density $n(\bm{r},t) $ for all eight density matrices defined in Fig.~\ref{latt}a) after the quench with Eq.~\eqref{quench} for $\theta=\pi/6$}
\label{evolution}
\end{figure}

\section{Tomography}
As we have shown, the dominant contribution to the local Chern marker comes from off-diagonal terms of nearest-neighbor and next-nearest-neighbor density matrices. These should be observable in experiments.
Higher-order contributions from far distant density matrices are in principle measurable as proposed in Ref.~\cite{Ardila2018}. For the measurement of the first-order contribution we propose a tomographic scheme \cite{Alba2011,Hauke2014,Flaschner2016,Ardila2018,Zheng2018} which measures nearest-neighbor and next-nearest-neighbor density matrices in real space and does not rely on an additional coupling channel as in Ref.~\cite{Ardila2018}. Our scheme can measure the eight different density matrices defined in Fig.~\ref{latt}a) individually. To this end, the system is quenched with the two-level Hamiltonian
\begin{equation}
H_Q=J\left(\cos\theta\sigma_x+\sin\theta\sigma_z\right).
\label{quench}
\end{equation}
The two levels  correspond to the two sites $\bm{r},\bm{r'}$ of the respective density matrix. The Hamiltonian in Eq.~\eqref{quench} corresponds to hopping between the sites with energy $J\cos\theta$ and an energy offset of $J\sin\theta$ between the levels. For $\theta=\pi/6$ the ratio between these energy scales is about 1.7. The quench can be performed by suddenly switching on a lattice potential which we schematically show in Fig.~\ref{evolution}a) for nearest-neighbor coupling and in b) for next-nearest-neighbor coupling. These potentials are created by superponing two retroreflected laser beams with wavevectors $\kappa$ and $2\kappa$. This yields a periodic double-well potential. The two lasers should exhibit a small phase difference in order to obtain an energy offset between the coupled sites. A third laser beam with wavevector $2\kappa$ determines the direction along which the two sites of the density matrix should be coupled. It points orthogonal for nearest-neighbor coupling and in 45$^\circ$ in next-nearest-neighbor coupling with respect to the direction of the aforementioned lasers. 

We parametrize the two-level density matrix as
\begin{equation}
\bar{\rho} = \left(\begin{matrix}
n(\bm{r}) & \rho(\bm{r},\bm{r'}) \\
\rho^*(\bm{r},\bm{r'}) & n(\bm{r'}) \\
\end{matrix}\right)
\label{para}
\end{equation}
where $n(\bm{r})$ is the on-site density of site $\bm{r}$. The time evolution of Eq.~\eqref{para} follows as
\begin{equation}
\bar{\rho}(t) = e^{iH_Qt}\bar{\rho} e^{-iH_Qt}.
\end{equation}
We find
\begin{equation}
\begin{split}
n(\bm{r},t) &= n(\bm{r})
 -  \left[n(\bm{r})-n(\bm{r'})\right]\cos^2(\theta)\sin^2(tJ)\\
&+\mathrm{Re}\rho(\bm{r},\bm{r'})\sin(2\theta)\sin^2(tJ)\\
&+\mathrm{Im}\rho(\bm{r},\bm{r'})\cos(\theta)\sin(2tJ).
\end{split}
\label{fit}
\end{equation}

By measuring the local densities $n(\bm{r},t) $ and $n(\bm{r'},t) $ as functions of time by means of a quantum gas microscope Eq.~\eqref{fit} can be used as a fit function to determine the off-diagonal part $\rho(\bm{r},\bm{r'})$ of the density matrix $\bar{\rho}$. In Fig.~\ref{evolution} we show the time evolution $n(\bm{r},t) $ for all eight density matrices defined in Fig.~\ref{latt}. The local density matrices can then be measured for the whole lattice at once. A measurement for $\theta=0$, i.e., without an energy offset, has been performed \cite{Atala2014} giving access to $\mathrm{Im}\rho(\bm{r},\bm{r'})$ in Eq.~\eqref{fit}. Artificial gauge fields in combination with a quantum gas microscope have made experimental progress in ladder systems \cite{Tai2017}.

\section{Conclusion}
In conclusion, we investigated the contributions to the local Chern marker in terms of off-diagonal density matrices for the instance of the two-band HHH model. We find that the first-order contribution is by orders of magnitudes the highest and purely local. Since topological properties are of course global properties, this gives only an indicator for topological non-trivial phases. At the topological phase transition the long-range correlation becomes large due to the gap closing.
We propose a tomographic measurement scheme for the first-order contribution which consists of measuring nearest-neighbor and next-nearest-neighbor correlations by means of a non-equilibrium superlattice quench and a quantum gas microscope. The two-band HHH model serves here as an example model. Applications to other models as well as extensions to multiple bands are straightforward. The generalization to the interacting case would require the many-body derivation of the original idea by Bianco and Resta \cite{Bianco2011} on many-mody Chern numbers \cite{Ishikawa1986,Wang2012}.

\begin{acknowledgments}
The authors would like to thank Monika Aidelsburger, Mathieu Barbier, Nigel R. Cooper, and Nathan Schine for enlightening discussions and Christian Schweizer for careful reading of the manuscript. This work was supported by the Deutsche Forschungsgemeinschaft (DFG, German Research Foundation) under Project No. 277974659 via Research Unit FOR 2414.
\end{acknowledgments}

\bibliographystyle{apsrev4-1}

%

\end{document}